\documentstyle[prl,aps,epsf,epsfig,multicol]{revtex}

\epsfclipon

\newcommand{\be}{\begin{equation}}
\newcommand{\ee}{\end{equation}}
\newcommand{\bes}{\begin{eqnarray}}
\newcommand{\ees}{\end{eqnarray}}
\newcommand{\bma}{\left( \begin {array}}
\newcommand{\ema}{\end {array} \right)}

\begin{document}

\title{Branched Polymers and Percolation}

\author{Peter Grassberger}

\address{John-von-Neumann Institute for Computing, Forschungszentrum J\"ulich, D-52425 J\"ulich, Germany}

\date{\today}

\maketitle
\begin{abstract}  
We study a supposed model for branched polymers which was shown in two dimensions 
to be in the universality class of ordinary percolation. We confirm this by high
statistics simulations and show that it is in the percolation universality class
also for three dimensions, in contrast to previous claims. These previous studies
seem to have been mislead by huge corrections to scaling in this model.

  \vspace{4pt}
\end{abstract}

\begin{multicols}{2}

Although most statistical physicists have an intuitive notion of the concept 
of universality at second order phase transitions, a rigorous method to delineate
universality classes does not yet exist. As a consequence, again and again the 
problem arises whether two specific models are in the same universality class or 
not. If critical exponents can be obtained exactly (either by conformal invariance 
in $d=2$, or by an $\epsilon$-expansion near the upper critical dimension), then 
one can at least say when they are not in the same class: if the exponents 
do not coincide. But in general one has to resort to numerical methods, such 
as series expansions or Monte Carlo simulations. Since both require extrapolations
in order to obtain critical exponents, it is not surprising that wrong claims
about universality classes appear again and again.

Several years ago, a model was introduced in \cite{lucena} (called LATST in the 
following) which was supposed to 
describe the growth of branched polymers in a disordered medium. Branched polymers
in thermal configurational equilibrium are in the universality class of lattice 
animals \cite{lubensky}. This is an ensemble of connected clusters of sites on 
a regular lattice where all configurations with the same number of sites 
have the same weight. 

The LATST model is different. It is defined kinetically. 
It lives on a randomly diluted hypercubic lattice, such that monomers can be 
placed only on a fraction $p$ of lattice sites.
Starting with a monomer as an active point seed, at each time step the oldest 
active site is chosen and its free neighbours (usable but not yet used) are 
counted. If there is no free neighbour, the site becomes inactive and the next 
active site in the list is chosen. Otherwise, one of the free neighbours (chosen 
randomly) is declared a new active site. In addition, if there is another free 
neighbour, a second site is made active with probability $b$. This is called 
`branching'. After this is 
done, the old active site is also made inactive and the next active site 
is chosen from the list. When $p$ is smaller than a critical value $p_c$, then
the process dies with probability 1 for all values of $b$. For $p>p_c$, there 
exist a critical value $b_c = b_c(p_c)$ such that it has a non-zero chance to 
survive for $b > b_c(p_c)$.

Obviously this is very similar to the growth of an epidemic in the general 
epidemic process (``dynamical percolation") \cite{grass}. In the latter, the 
main difference is that all neighbours of an active site have the same chance 
$p$ to be activated, independent of how many other can be or are activated.
Thus, if a site has $m$ neighbours not yet occupied, 
the number of its ``descendants" is a Poissonian random number with
average $p\,m$. In contrast, in the LATST model the variance of the number 
of activated descendants is reduced. It is always either 0, 1, or 2, but never 
$>2$. This difference is most pronounced for $p=1$ and for lattices with 
high coordination number, because then $b_c$ is very small (see below) and 
most of the time there is just a single descendent. 
But we should not expect this to be relevant in the 
renormalization group sense, i.e. we should expect that the LATST model is in 
the universality class of the general epidemic process and hence of ordinary 
percolation. In some sense, LATST and general epidemic process are 
related to each other like `growing self avoiding walk' (GSAW) 
\cite{majid,lyklema} and usual self avoiding walk (SAW) which are also known to 
be in one common universality class. 

The transition in the 2-dimensional LATST model was indeed found to be in the 
percolation universality class in \cite{porto}. For $d=3$, the same was 
verified in \cite{porto} for values of $p$ close to $p_c$, but not for $p=1$. 
In \cite{aragao} it was claimed that the latter shows clean scaling which is 
definitely not in the percolation universality class. We claim here that this 
is wrong. With present day computers it is practically impossible to obtain
the scaling regime for the case $d=3, p=1$, but all numerical evidence hints 
to the fact that the model is in the percolation universality class.

In order to see the origin of the problem, we notice that the LATST model is
for $p=1$ and $b=0$ just the GSAW. As already pointed out, this is in the SAW 
universality class. But it has greatly reduced attrition since monomers are 
placed only on free neighbours in the GSAW, while they are placed at 
randomly chosen neighbours (whether they are free or not) in SAWs. As a 
consequence, the attrition constant (the rate with which the process dies) 
is 0.024 in $d=2$ and 0.000275 in $d=3$. Thus, in order to overcome attrition, 
it would be sufficient to make an enrichment step \cite{wall} every 40 time 
steps in $d=2$, and every 3600 time steps in $d=3$. These give lower estimates
$b_c(p=1) \ge 0.024 \;(d=2)$ and $b_c(p=1) \ge 0.000275 \;(d=3)$. Indeed, the 
actual estimates for $b_c$ are rather close to these: $b_c(p=1) = 0.056$ 
\cite{porto} for $d=2$ and $b_c(p=1) = 0.00034$ \cite{aragao} for $d=3$.
Since $1/b$ sets a time scale (the average time between two branchings), 
this shows that there are large inherent times $T \approx 20$ resp. 
$T\approx 3000$ in the LATST model. Any critical scaling is expected to 
show up only for $t \gg T$. While such $t$ are still feasible in high-statistics
simulations in 2d, they are out of reach in 3d.

To verify these predictions, we performed simulations. We studied only the 
most difficult and controversial case $p=1$, for $p<1$ it seems accepted
that the LATST model is in the percolation class. In order to reach large
clusters without finite lattice effects we used hashing. This allowed us 
to use virtual lattices of sizes up to $100,000^3$, so we could study 
clusters of up to $10^6$ sites without any finite lattice corrections.
Such corrections were obviously important in the figures shown in \cite{aragao},
but it is hard to judge from these figures what parts of the distributions are
uneffected by them.

We used the fast and reliable 4-tap random number generator of \cite{ziff}.
Each curve in the following figures is based on a sample with at least
$3\times 10^5$ clusters. Although this statistics is higher than previous 
ones \cite{lucena,porto,aragao} by several orders of magnitude, the entire 
project needed only about 200h CPU time on fast work stations. 

Mass distributions of 2-d clusters for 5 values of $b$ are shown in Fig.1.
Actually, in order to take into account the very small error bars, we 
plotted not $P(M)$ itself but $M^{\tau-2} P(M)$ where $P(M)$ is the 
probability that a cluster has mass $\ge M$, and $\tau=96/91$. This is 
motivated by the fact that $P(M) \sim M^{2-\tau}$ for 2-d percolation. Thus
we expect our data to be flat for $b=b_c$, except for corrections to scaling.
This is indeed the case for $M > 10^4$ and $b_c = 0.05680\pm 0.00004$. The 
average square size $\langle R^2\rangle$ obtained from the sdame runs is 
shown in Fig.2. Again the asymptotic behaviour $R^2\sim M^{96/91}$ expected 
for percolation is divided out, and again we see flat curves for $M > 10^4$.
Our estimate for $b_c$ is in rough agreement with that in \cite{porto} but 
about ten times more precise (assuming that the error quoted on page 1744 of 
\cite{porto} is misprinted; otherwise it would be a factor 100 more precise).

For $d=3$, $M^{\tau-2} P(M)$ versus $M$ is plotted in Fig.3. For $\tau-2$
we used the value $0.189$ from \cite{ziff}. We see clearly different 
behaviour for $M<3000$ and for $M>3000$. For $M<3000$ we have the expected
scaling of simple random walks. For $M\gg 3000$ we expect to
see percolation, but this sets in only very late. Roughly straight lines 
in the range $3000 < 10^5$ are seen for $b\approx 0.00034$. Obviously, 
the estimate
$b_c=0.000334$ of \cite{aragao} was based on this range, although
all log-log plots of \cite{aragao} show straight lines in the 
entire range $M > 10^2$ for reasons which we do not understand. Anyhow, 
these straight lines do not represent the 
asymptotic behaviour, since all our curves except those for $b \ge 0.00036$
bend down for very large $M$ 
(notice that each curve is based on independent runs, thus all systematic
structures seen in any of our plots are significant). Indeed we estimate that
the curve for $b=0.00036$ in Fig.3 will also bend down for $M > 10^6$, and the 
estimated critical value is $b_c = 0.000366\pm 0.000004$. 

\begin{figure}[b]
  \begin{center}
    \psfig{file=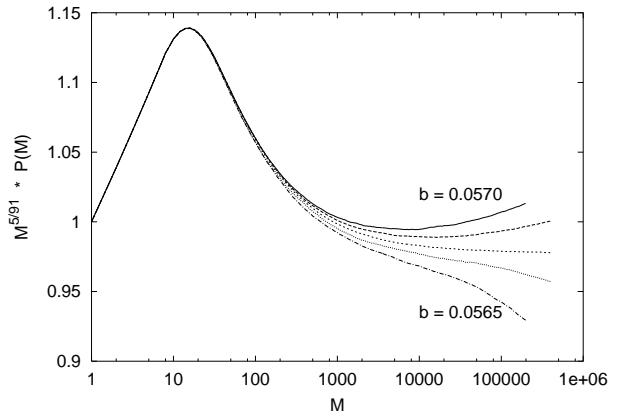,width=5.8cm,angle=270}
    \vglue0.2cm
    \begin{minipage}{8.5cm}
      \caption{Log-linear plot of $M^{\tau-2} P(M)$ versus $M$ for 2-d clusters.
       The lines are for $b=0.057, 0.0569, 0.0568, 0.0567$, and $0.0565$ (top to 
       bottom).
      }
  \end{minipage}
\end{center}
\label{fig1}
\end{figure}

\begin{figure}[b]
  \begin{center}
    \psfig{file=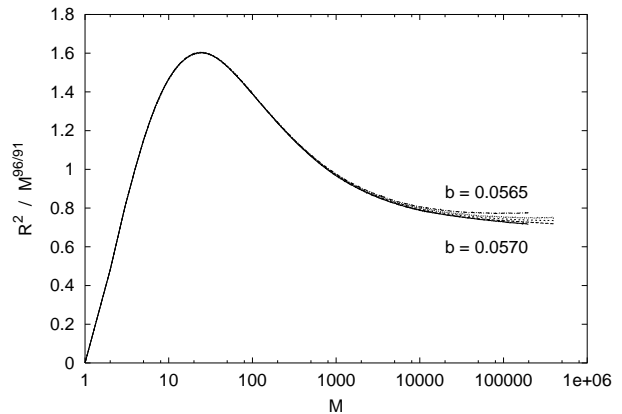,width=5.8cm,angle=270}
    \vglue0.2cm
    \begin{minipage}{8.5cm}
      \caption{Log-linear plot of $R^2 / M^{2/D_F}$ versus $M$ for 2-d clusters,
       where $D_F=91/48$ is the fractal dimension of 2-d percolation clusters.
       The curves correspond to the same $b$-values as in Fig.1 (bottom to top).
      }
  \end{minipage}
\end{center}
\label{fig2}
\end{figure}

Average values of the ``chemical radius" (the length of the paths connecting 
active sites to the seed) and of 
the number of active sites are shown in Figs. 4 and 5. Here we plotted the raw 
data themselves. We again see the break in scaling at $M\approx 3000$. A 
careful look at the data (e.g. plotting them again with some power split 
off) shows that the curves are not straight for $M\gg 3000$ but show similar 
curvature as $P(M)$. Thus also in these cases it is impossible to extract critical 
exponents from 
the data. In contrast to this, the authors of \cite{aragao} found 
perfect scaling without visible corrections (and with 
exponents different 
from those for percolation) for all cluster masses $>10^2$ and chemical radii $>10$.
We have no explanation for this.

In summary we showed that it is extremely dangerous to estimate critical 
exponents from data which have important corrections to scaling. In the present 
case these corrections seem to have lead to wrong claims about universality, 
but it is not clear why the authors of \cite{aragao} have missed them. In any 
case we see no reason to doubt that the LATST model is in the universality 
class of percolation, for any finite dimension $d$. Numerical verification for 
$d\ge 4$ is of course out of question.

\begin{figure}[b]
  \begin{center}
    \psfig{file=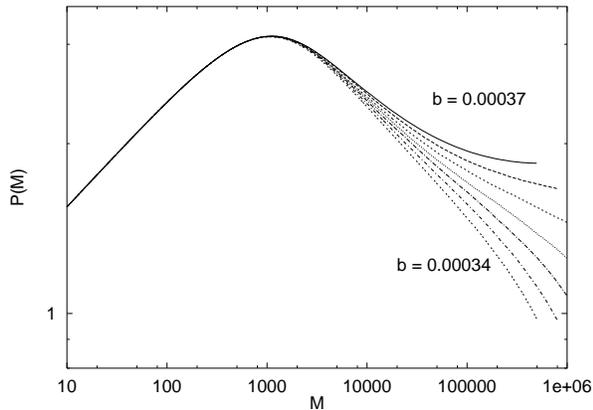,width=5.8cm,angle=270}
    \vglue0.2cm
    \begin{minipage}{8.5cm}
      \caption{Plot of $M^{0.189} P(M)$ versus $M$ for 3-d clusters.
       The lines are for $b=0.00037,0.000365,0.00036,\ldots ,0.00034$ 
       (top to bottom).
      }
  \end{minipage}
\end{center}
\label{fig3}
\end{figure}

\begin{figure}[b]
  \begin{center}
    \psfig{file=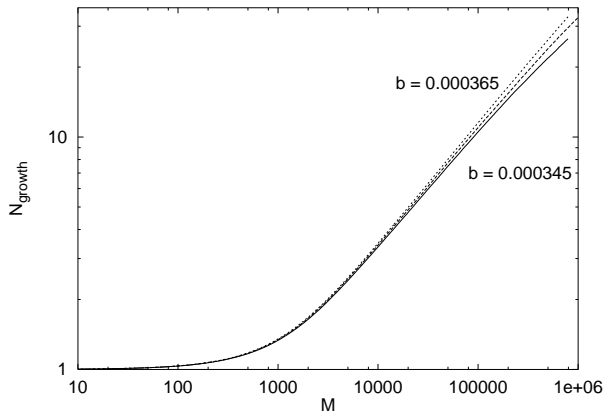,width=5.8cm,angle=270}
    \vglue0.2cm
    \begin{minipage}{8.5cm}
      \caption{Number of active (growth) sites versus $M$ for 3-d clusters.
       Only three curves are shown.
      }
  \end{minipage}
\end{center}
\label{fig4}
\end{figure}

\begin{figure}[b]
  \begin{center}
    \psfig{file=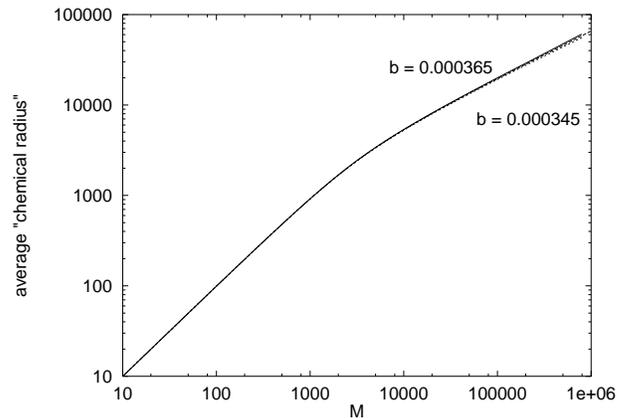,width=5.8cm,angle=270}
    \vglue0.2cm
    \begin{minipage}{8.5cm}
      \caption{``Chemical radius" (average shortest path length from seed to 
       active sites) versus $M$ for 3-d clusters.  Only three curves are shown.
      }
  \end{minipage}
\end{center}
\label{fig5}
\end{figure}

Let us finally end with two comments. The first is that the subcritical 
process is not, as claimed in \cite{lucena,porto,aragao}, in the universality 
class of SAWs. It is in the lattice animals universality class, like
any model whose critical 
\linebreak[4]
\noindent
point is in the percolation class \cite{footnote}. Secondly, it is not 
clear in what sense the LATST model is a valid model for branched polymers (BP).
It certainly does not describe the ensemble of fully annealed BP, except in 
the sense just mentioned (lattice animals are in the same universality class as 
BP). It cannot describe either the ensemble of BP in which branch points are 
frozen but bends between successive monomers are annealed (i.e., BP with fixed 
topology), since this topology is determined in the LATST model by 
local configurations. It is for that reason that we avoided the name `branched 
polymer growth model' (BPGM) proposed in \cite{porto,aragao}.

\end{multicols}


\begin{references}
\bibitem{footnote} This is at least true if one conditions on rare large clusters 
   with fixed number of sites, as is usually done. If one conditions instead 
   on clusters 
   with large fixed chemical radius, then subcritical percolation (and thus also 
   LATST) is in the SAW universality class.
\end{references}
\end{document}